# Soliton disentangling and ferroelectric hysteresis in reconstructed moiré superlattices

Yanshuang Li[1,2†], Huan Zeng[1,2†], Xiuhua Xie[1*], Binghui Li[1], Jishan Liu[3], Shuangpeng Wang[4], Dengyang Guo[5], Yuanzheng Li[6], Weizhen Liu[6], Dezhen Shen[1*]

[1]State Key Laboratory of Luminescence and Applications, Changchun Institute of Optics, Fine Mechanics and Physics, Chinese Academy of Sciences, No. 3888 Dongnanhu Road, Changchun, 130033, China

[2]University of Chinese Academy of Sciences, Beijing 100049, China

[3]Center for Excellence in Superconducting Electronics, State Key Laboratory of Functional Materials for Informatics, Shanghai Institute of Microsystem and Information Technology, Chinese Academy of Sciences, Shanghai 200050, China

[4]MOE Joint Key Laboratory, Institute of Applied Physics and Materials Engineering and Department of Physics and Chemistry, Faculty of Science and Technology, University of Macau, Macao SAR 999078, P. R. China

[5]Department of Physics, Cavendish Laboratory, University of Cambridge, Cambridge, UK

[6]Key Laboratory of UV-Emitting Materials and Technology, Ministry of Education, Northeast Normal University, Changchun 130024, China

*Corresponding authors. Email: xiexh@ciomp.ac.cn; shendz@ciomp.ac.cn

†Yanshuang Li and Huan Zeng contributed equally to this work.

## ABSTRACT

Moiré materials, created by lattice-mismatch or/and twist-angle, have spurred great interest in excavating novel quantum phases of matter. Latterly, emergent interfacial ferroelectricity has been surprisingly found in spatial inversion symmetry broken moiré systems, such as rhombohedral-stacked bilayer transition metal dichalcogenides (TMDs). However, the evolution of moiré superlattices corresponding to polarization switching and hysteresis is still unclear, which is crucial for giving insight into the interplay between lattice symmetry and band topology, as well as developing optoelectronic memory devices. Here we report on the observation of phonon splitting at strain soliton networks in reconstructed moiré superlattices, arising from the twisting and relaxing induced strong three-fold rotational symmetry ($C_3$) breaking. The interval of phonon splitting is tunable by a perpendicular displacement field and exhibits ferroelectric-related hysteresis loops. These phonon evolution features are attributed to the contribution of moiré solitons disentangling and lattice viscosity during the motion of domain walls. Moreover, we demonstrate a proof-of-principle moiré ferroelectric tunneling junction, whose barrier is modified by net polarization with a tunneling electroresistance of ~$10^4$. Our work not only reveals the lattice dynamics of moiré solitons but also presents a potential pathway for future ferroelectric-based optoelectronic memory devices.

**INTRODUCTION**

Moiré superlattices, arising from a mismatch in lattice constant or orientation between stacked monolayers, have emerged as powerful platforms for uncovering intriguing quantum states of matter, such as superconductivity[1,2], correlated insulator[3,4], orbital magnetism[5,6], etc. With these inspiring advances, one of the primary concerns of moiré physics is becoming exploring emergent phenomena which absent from the individual constituents. For example, the long-period superlattices can fold the energy band into the moiré mini-Brillouin zone[7], giving rise to a flat band and moiré potential[8,9], altering electronic correlations and topology, which is unattainable in the pristine band. In addition, engineering the symmetry breaking in moiré systems also offers multi-faceted approaches to tuning the internal quantum degrees of freedom[10,11]. Recently, unconventional interfacial ferroelectricity has been found in lattice inversion symmetry broken homo-bilayer structures, such as parallel-stacked bilayer hexagonal boron nitride (hBN)[12,13], and rhombohedral-stacked (R-stacked) bilayer transition metal dichalcogenides (TMDs)[14,15]. The out-of-plane electric dipole moment stems from the local atomic registry and interlayer hybridization, switchable by in-plane interface sliding[16-18]. This kind of emergent ferroelectricity incorporate in semiconducting R-stacked bilayer TMDs, namely bilayer $MoS_2$, $WS_2$, $MoSe_2$, and $WSe_2$, show a potentiality for developing non-volatile memory devices[19,20]. Moreover, combined with unique excitons properties (spin-valley locking, gate-tunable nontrivial topological phase transition) and interface polarization in R-stacked bilayer TMDs, inspire interest in future intelligent optoelectronic device applications[21].

In marginally twisted R-stacked TMDs homo-bilayers, real moiré superlattices undergo lattice relaxation as a result of competition between intralayer strains and interlayer van der Waals (vdWs) coupling, leading to the reconstruction of energetically preferential domains with staggered stacking order[22-25], which differ from rigidly twisted lattices (Fig. 1a, b). Commensurate domains possess alternating polarization, separated by a network of strain twist-faults, termed moiré solitons[23,26,27], acting as domain walls (DW) (Fig. 1c). The contrary evolution of adjacent domains under an out-of-plane electric displacement field has been proposed by theoretical works[18,28] and shown by microscopy experiments[14,15]. Although some hysteresis behavior of net polarization has been discovered, a microscopic sight of DW dynamics at the lattice level, which is fundamental for the ferroelectric response, remains inconclusive. Moreover, studies of resistive switching induced by moiré ferroelectricity are also insufficient.

Herein we reveal a notable correlation between the strain of moiré solitons and polarization hysteresis. At the soliton networks, local strain distorts lattice and breaks the three-fold rotational symmetry ($C_3$), causing the corresponding Raman vibrational mode ($E_{2g}$) to split. The out-of-plane electric field dependent Raman frequency spacing allows us to identify DW motion, namely, the process of moiré solitons disentangling and entangling. Hence, we interpret the micro-mechanism of moiré ferroelectric hysteresis by using the interlayer lattice viscosity observed during moiré soliton disentanglement. Additionally, we demonstrate proof-of-principle moiré ferroelectric tunneling junctions to better elucidate the real function of net polar domains for the prospective applications of nonvolatile switching. In light of these new findings, moiré superlattices with inherent ferroelectricity thus likely not only offer a deeper insight into the strong correlation and topological physics recently observed in twisted bilayer TMDs, but also put forward promising scenarios for the novel electronic and optoelectronic devices with memory functions.

## RESULTS AND DISCUSSION

**Phonon Splitting in Moiré Solitons Networks.**

In the range of small twisting angles of homo-bilayers, as a consequence of the balance between intralayer and interlayer energies, the rigid atomic lattice suffers straining, along with moiré superlattice reconstructions. Quasi-one dimensional moiré solitons, formed between polar domains, possess the information on the interlayer twist angle, which induces non-uniform uniaxial strain. The value of local strain on solitons is dependent on the twisting angle and degree of moiré superlattice reconstruction. In order to precisely control the twisting angle, we utilized the hBN-assisted tear-and-stack method[29] to make a set of twisted bilayers $MoS_2$, with twisting angles from 0° to 8°, which means realizing the R-stacked structure, breaking global inversion symmetry and the out-of-plane mirror symmetry, concurrently (more details about the sample preparation procedure can be found in Methods and Supplementary Information). We performed Raman spectroscopy measurements that allow the identification of vibrational phonon modes in the reconstructed moiré superlattice, modified by the strain that originates from interlayer lattice relaxation. The angle evolution behavior of Raman phonons peak positions of the twisted samples we studied is illustrated in Fig. 2a, with data measured in the high-frequency range, exhibited two typical intralayer vibration modes, known as $E_{2g}$ and $A_{1g}$. Notably, the originally energy degenerate $E_{2g}$ phonon vibrational mode is split into two peaks within a specific twisting angle range. Considering that the $E_{2g}$ mode is caused by the opposing movements of two sulfur (S) atoms with respect to the molybdenum (Mo) atom in the in-plane direction (sketched in Fig. 2b right), phonon splitting indicates that $C_3$ symmetry has been broken by uniaxial strain. A slight shift of $A_{1g}$ mode with twisting angle varying may be a result of the three-dimensional reconstruction of superlattices[30], in which the out-of-plane relative motions of the S atoms can be modified. It also can be seen that the splitting of the E$_{2g}$ mode is divided into three classes within the twisting angle range we studied, viz. from unimodality to bimodality, and reverted to the initial state finally, which signifies the degree of moiré superlattice reconstruction[31].

As illustrated in Fig. 2b left, between two mirror-symmetrical commensurate domains, moiré soliton, arising from interlayer lattice entanglement, endures significantly uniaxial strain alone in the armchair direction, due to lattice relaxing. Twisting and reconstruction induced lattice distortions to make part of the Mo-S bond stiff and part soften, leading to $E_{2g}^{+}$ and $E_{2g}^{-}$ arising blueshift and redshift, respectively, relative to initial E$_{2g}$, as shown in Fig. 2c. Therefore, the spacing between $E_{2g}^{+}$ and $E_{2g}^{-}$ reflects the degree of distortion of the soliton lattice. Likewise, strain reduction is also manifested in a reduction in spacing, see further discussion below.

Figure 2d shows more specific signatures of E$_{2g}$ phonon splitting at different twist angles, from 0° to 8°, extracted from Fig. 2a. The largest splitting up to 3.4 cm$^{-1}$ around 3.4°, which suggests the greatest strain. At the small twisted angle (0°<θ<2.8°), although the moiré superlattice experienced a reconstruction, a minor twist fault in the soliton makes the lattice distortion insufficient, which is not enough to split the E$_{2g}$ mode. Correspondingly, at the larger twist angle (6°<θ<8°), when the interlayer vdWs energy is not enough to offset the strain energy in the intralayer, the lattice relationship of the bilayer returns to a rigid state of monolayers, that is, the area of incommensurate atomic registries is increased instead, and the lattice distortion is not obvious. Remarkably, at the range of 2.8°<θ<6°, the E$_{2g}$ phonon mode splits into two distinct peaks, indicating moiré solitons possess high strain locally, as a consequence of the combination of twisting and relaxing, distorted lattice breaking the $C_3$ symmetry.

**Electrically Tunable Moiré Solitons Disentangling.**

Since adjacent commensurate domains keep opposite dipole moments, an external electric field with an out-of-plane direction will drive an increase in the area of domains aligned with its own polarization[14]. The area evolution of domains also implies the movement of the DW, that is, the movement of lattice solitons in the reconstructed moiré superlattice. Considering that the soliton movement will inevitably cause the change of strain, we employed electric field-dependent Raman spectroscopy to study the strain variation of soliton, where a dual-gate structure (Fig. 3a) was fabricated to control the perpendicular displacement field. Notably, the twisting angle of R-stacked bilayer $MoS_2$, in this case, was fixed in 3°, for a higher strain, as we observed above, which is expected to have more pronounced Raman peaks shifting varying with the electrical field.

Figure 3b shows the tendency of high-frequency phonon mode ($E_{2g}$ and $A_{1g}$) as a function of the external electric field. With the gate voltage increasing, the peak interval of $E_{2g}^+$ and $E_{2g}^-$ gradually diminish (Fig. 3b, c), while the peak of $A_{1g}$ is stationary, indicating that the out-of-plane electric field affects the distorted lattices of solitons and tends to restore it to $C_3$ symmetry. Interestingly, as shown in Fig. 3c, the $E_{2g}^+$ manifest an apparent redshift compared to $E_{2g}^-$, which suggests the stiffness Mo-S bond of solitons becoming softened. The redshift of the peak position of $E_{2g}^+$ tends to plateau after 20 V, which indicates that moiré solitons lattice strain driven by the perpendicular electric field reaches a new equilibrium. These electric field-dependent changes in the Raman phonon vibrational modes illustrate the dynamic evolution of solitons lattice, which can be phenomenologically illustrated in Fig. 3e. When no gate voltage is applied, the soliton lattice is entangled and hosts large lattice distortion along the armchair direction (middle of Fig. 3e), so that the stacking order of the bilayer favorably transitions from MX (top metal atoms, bottom chalcogen atoms) to XM (top chalcogen atoms, bottom metal atoms). With the positive electric field applied, as shown at the top of Fig. 3e, the MX domain aligned with the polarization direction of the external electric field will gradually dominate, and the corresponding moiré soliton will squeeze the space of the XM. The movement of the moiré soliton appears as a disentangling process of the strained lattice, which is therefore accompanied by a tendency to restore $C_3$ symmetry. The bottom of Fig. 3e illustrates the opposite situation.

To further characterize the hysteresis behavior of the movement of the moiré soliton, by applying a closed-loop voltage (inset of Fig. 3d), the $E_{2g}$ Raman frequency spacing vs voltage ($\Delta\omega$-V) measurement was performed in this dual-gate structure. The $\Delta\omega$ in Fig. 3d shows a butterfly-shaped hysteresis loop, indicating the ferroelectricity presented. The attenuation of $\Delta\omega$ with the applied electric field marks the disentangling process of the soliton, however, the existence of the coercive field means that the interlayer lattice viscosity of the soliton plays a role in the re-entangling process. Therefore, solitons disentangling and lattice viscosity results in the switching of moiré ferroelectricity, as illustrated in Fig. 3e.

**Ferroelectric Tunneling and Resistive Switching.**

The working mechanism of moiré ferroelectricity is fundamentally different from a conventional ferroelectric. As we discussed above, moiré ferroelectricity hosts two opposite polar domains in a reconstructed moiré unit cell. Thus, the overall interfacial barrier modulation induced by interlayer polarization in moiré superlattices will exhibit unique resistance-switching properties. Here, we fabricated a Graphene/0°-R-stacked bi-$MoS_2$/hBN/Graphene structure, a two-terminal vertical moiré ferroelectric tunnel junction (MFTJ), as shown in the upper left inset of Fig. 4a. Current-voltage (I-V) measurement has been performed, where the bias voltage is pulse mode (pulse duration, 200 ms) for reducing thermal effects, as shown in the lower right inset of Fig. 4a. The corresponding tunneling current of the MFTJ presents a typical ON/OFF and hysteresis features above ~±5 V bias, suggesting the change of net moiré ferroelectric polarization. Remarkably, at a smaller bias voltage range, approximately from -5 V to +5 V, MFTJ exhibits a high resistance state. A reasonable explanation is that an inadequate perpendicular electric field could not make enough area difference of opposite polar domains (between MX and XM), generating insufficient interface barrier modification for current tunneling. As the bias voltage increases, one of the polar domains gradually becomes dominant, which in turn puts the device in a low resistance state. Subsequently, when the applied voltage is reduced, the existence of interlayer lattice viscosity tends to maintain the lattice status and produce moiré ferroelectric hysteresis. The asymmetric curves in the positive and negative bias range of the MFTJ may be attributed to the global strain and the asymmetry of the domains caused by the non-uniform twisted angle. Zoom in to see I-V curves from the positive bias range (right of Fig. 4a), the switch ratio of the MFTJ is up to ~$10^4$ at room temperature. It should be noted that here the compliance current was set at 100 nA to prevent device breakdown. Thus, the device switch ratio above should be underestimated. Furthermore, for comparison purposes, we also fabricated the H-stacking bilayer $MoS_2$ tunneling junctions, which have no ferroelectric hysteresis behavior (Supplementary Figure S6). Therefore, this confirms that the resistive state switching in the device originates from the switching in the moiré ferroelectric polarization.

For further study of the modified interface potential barrier, the Fermi-level shift in graphene was confirmed by employing Raman spectroscopy measurements. As shown in Fig. 4b, the G and 2D modes of the graphene contacted with R-stacked bilayer $MoS_2$ both have evident blueshifts compared to the isolated graphene region. The main reason for these blueshifts is the fact that the polarization field of the commensurate domains changes the doping level of the graphene. The Raman shift of the G-mode peak ($\Delta\omega$) is about 6.06 cm$^{-1}$, which can be used to quantitate the shift of graphene Fermi level ($\Delta E_f$), $|\Delta E_f| \approx \frac{\Delta\omega}{21} \approx 0.29 eV$[32,33]. During blue-shifting, the intensity ratio between the 2D and G modes ($I_{2D}/I_G$) decreased, which is the result of the doping effect by the polarization electric field[34,35]. Considering the reconstructed R-stacked bilayer MoS$_2$ have two opposite polar domains, as mentioned above, graphene can be doped in electron- and hole-type alternatively. Therefore, a large proportion of graphene will be depleted, resulting in an underestimate of Fermi level shifts estimated by G-mode Raman shifts. This is also in consonance with the high resistance state of the I-V curve in the small bias voltage range, where there is not enough net polar domain to change the interface potential barrier, further verifying the moiré ferroelectricity feature.

The band diagrams and operational principle of our MFTJ are shown in Fig. 4c. The tunnel barrier is consist of bi-MoS$_2$ and hBN, with enough height/width, to allow nearly no thermionic emission and tunneling current at low bias voltage, namely OFF state. As the positive bias voltage increases, the perpendicular displacement field makes the polar domain (commensurate XM region) become dominated, associated with a net polarization field emerged, whose effect is to shift the Fermi level of graphene from the Dirac point to n-type (the right side of Fig. 4c). The extra shifting of the Fermi level in graphene decreases the average barrier height and width, inducing a larger tunneling current. Under reverse bias, a relatively symmetrical situation occurs (no schematic here).

# CONCLUSION

We demonstrated the moiré solitons dynamics signatures and ferroelectric hysteresis in reconstructed moiré superlattices. The intrinsic lattice distortion in soliton networks is derived through the $E_{2g}$ mode splitting in Raman spectroscopy. The shrinking of the phonon splitting under the penetration electric field suggests that the strain-broken $C_3$ symmetry tends to recover, indicating the occurrence of moiré solitons disentanglement process. Then, the lattice viscosity has been verified by the coercive field that exists in the hysteresis of $E_{2g}$ Raman frequency spacing. The MFTJ based on Graphene/0°-R-stacked bi-$MoS_2$/hBN/Graphene vdWs heterostructure exhibit a high ratio of tunneling current, nearly $10^4$. Unlike conventional ferroelectric switching, our device relies on the competition between polarized domains (XM/MX) to switch the net polarized electric field. Looking ahead, our findings point to many intriguing possibilities. For instance, the evolution of the soliton network and polarized domains may induce the topological phase transition of the energy band, thereby affecting the quantum geometry of the Bloch wave function, which is vitally manipulating the ground/excited state behavior of quasiparticles[21,36,37]. Moreover, moiré ferroelectricity also provides a new approach for in-memory computing (beyond von Neumann architecture) [38,39].

# METHODS

**Device fabrication.** Monolayer $MoS_2$, graphene, and multilayer hBN are made by Scotch-tape mechanical exfoliation method from $MoS_2$ crystals (Six Carbon Technology, Shenzhen), natural graphite (NGS, Germany), and hBN crystals (HQ graphene, Netherlands). All homo- and hetero-stacking devices used the hBN-assisted 'Tear-and-stack' method (See the video in supplementary information). Firstly, we put polypropylene carbonate (PPC) dissolving in chloroform solution (6% weight) and scribble it on the surface of polydimethylsiloxane (PDMS). The PDMS/PPC is the stamp to pick up all materials. Secondly, we picked up the top hBN to tear monolayer $MoS_2$ via the stamp. Then, we rotated the other $MoS_2$ on the substrate and picked it up. Thirdly, we put these on above of bottom hBN and heated them to 90 °C for 5 minutes. Finally, we used chloroform, acetone, and isopropanol to dissolve and clean PPC and used the $N_2$ flow to dry it. The Au electrodes are pre-lithographically etched on the substrate by a standard lithography process.

**Optical and morphology characterization.** The optical images were obtained in a homemade microscopy system (20× objective lens). Raman spectra were measured by the Horiba-LabRAM-HR-Evolution instrument (Horiba, Japan) at room temperature. The emission wavelength of the laser source is 532 nm and the maximum emission power is about 8.7 mW. The laser spot is about 1 um vis a 100× objective lens (Olympus Corporation, Japan) focusing. The height of hBN was measured by atomic force microscope (Multimode8, Bruck Instruments, Germany).

**Current–voltage measurements.** I-V measurements were performed by the semiconductor analyzer instrument (B1500A, Keysight Technologies, America) at room temperature. The bilayer $MoS_2$ side was connected to the ground.

**ACKNOWLEDGMENTS**

We gratefully acknowledge X. Z. Yan and Z. Yu for their assistance with Raman measurements and fruitful discussions. This research is mainly supported by the National Natural Science Foundation of China (Grant Nos. 11727902). S.P. and X.H. gratefully acknowledge support from Multi-Year Research Grants (MYRG2020-00207-IAPME) from Research & Development Office at University of Macau.

**Competing interests**

The authors declare no competing interests.

**Data Availability**

All data that support the findings of this study within this paper and the Supplemental Information are available from the corresponding authors upon reasonable request.

**Figure Captions**

**Figure 1. Schematic illustration of lattice reconstruction and the existence of polarization in moiré superlattice.** (a) Moiré superlattices present different lattice structures in a rigid and relaxed situation. (b) The origin of polarization in R-stacked bilayer $MoS_2$ (the yellow and blue ball is the sulfur atom and molybdenum atom, respectively). (c) Polarization direction and domain wall in reconstructed moiré superlattices.

**Figure 2. Raman spectra of R-stacked bi-$MoS_2$ and analysis.** (a) Measured Raman spectra of bilayer $MoS_2$ for different twist angles. (b) The model of domain wall and intralayer phonon mode (the yellow and blue ball is the sulfur atom and molybdenum atom, respectively); the $E_{2g}$ mode related to in-plane motion between molybdenum and sulfur atoms. (c) The blue line is the Raman spectrum of $E_{2g}$ mode at the 3° twist angle, the red line is the fitted curve. (d) Specific signatures of $E_{2g}$ phonon splitting at different twist angles.

**Figure 3. The electric field-dependent Raman spectroscopy, soliton disentangling, and lattice viscosity.** (a) Schematic illustration of bilayer twisted $MoS_2$ device. (b) Raman spectra under different out-of-plane electric fields in the wavenumber range of $E_{2g}$ and $A_{1g}$ modes. The excitation laser is 532 nm. (c) The fitting peaks of $E_{2g}^{-}$ and $E_{2g}^{+}$ vs bias voltages. (d) The hysteresis curve of Raman frequency spacing of $E_{2g}$, suggests the lattice viscosity effect. The applied voltages are shown in the inset (right-top). (d) Schematic illustration of moiré soliton disentangling under the out-of-plane electric field.

**Figure 4. The moiré ferroelectric tunnel junction based on R-stacked bilayer $MoS_2$.** (a) Current-voltage curves of the moiré ferroelectric device. Insert schematic illustration is the structure of the device (left-top). Bilayer $MoS_2$ side is grounded. The applied bias voltages are shown in the inset (right-bottom). (b) The Raman spectra of graphene with/without the R-stacked bilayer $MoS_2$. The excitation laser is 532 nm. (c) Energy band schematic illustration for the OFF and ON states of our ferroelectric device.

**Figure-1**

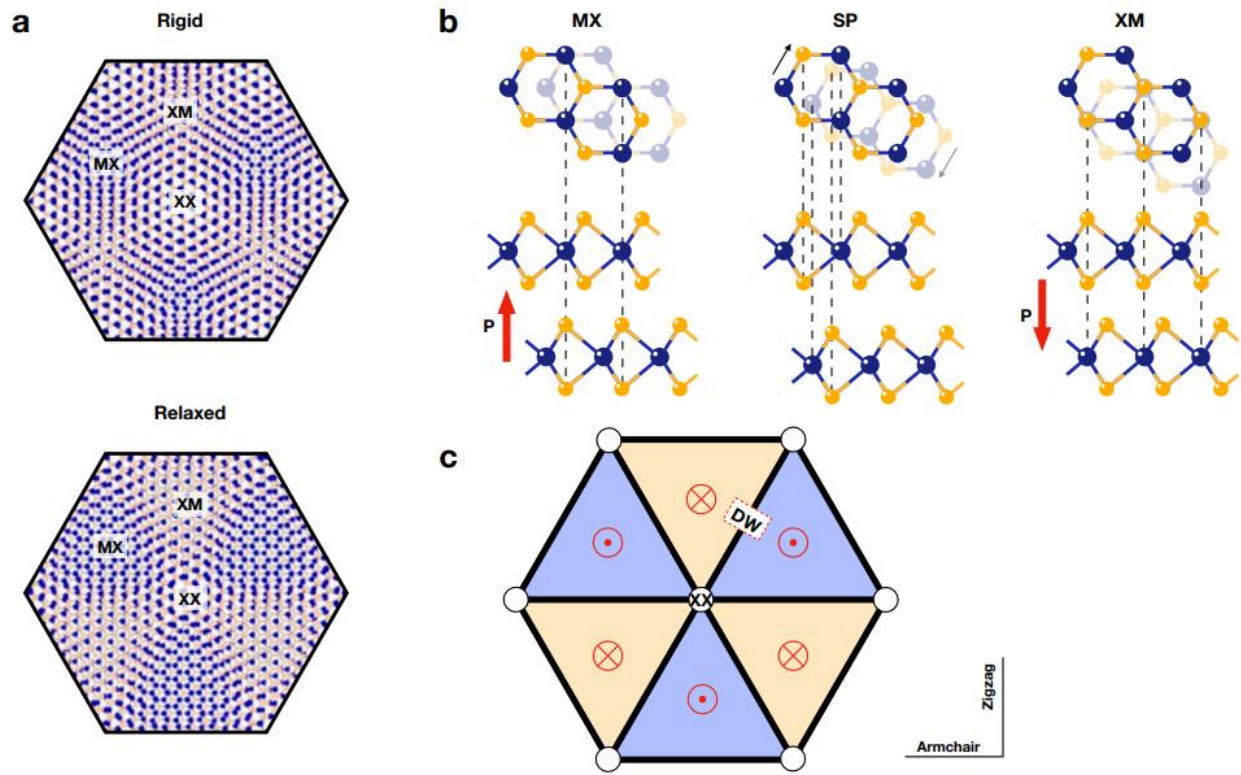

**Figure-2**

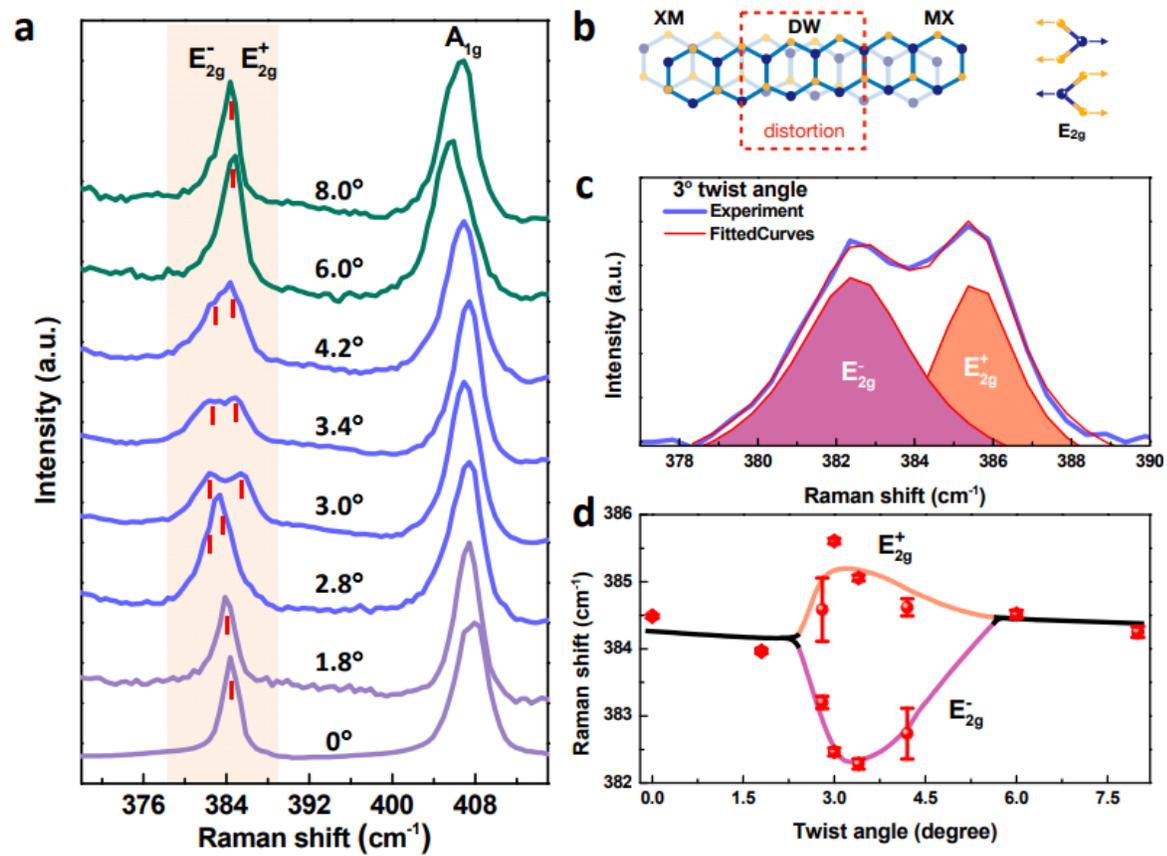

**Figure-3**

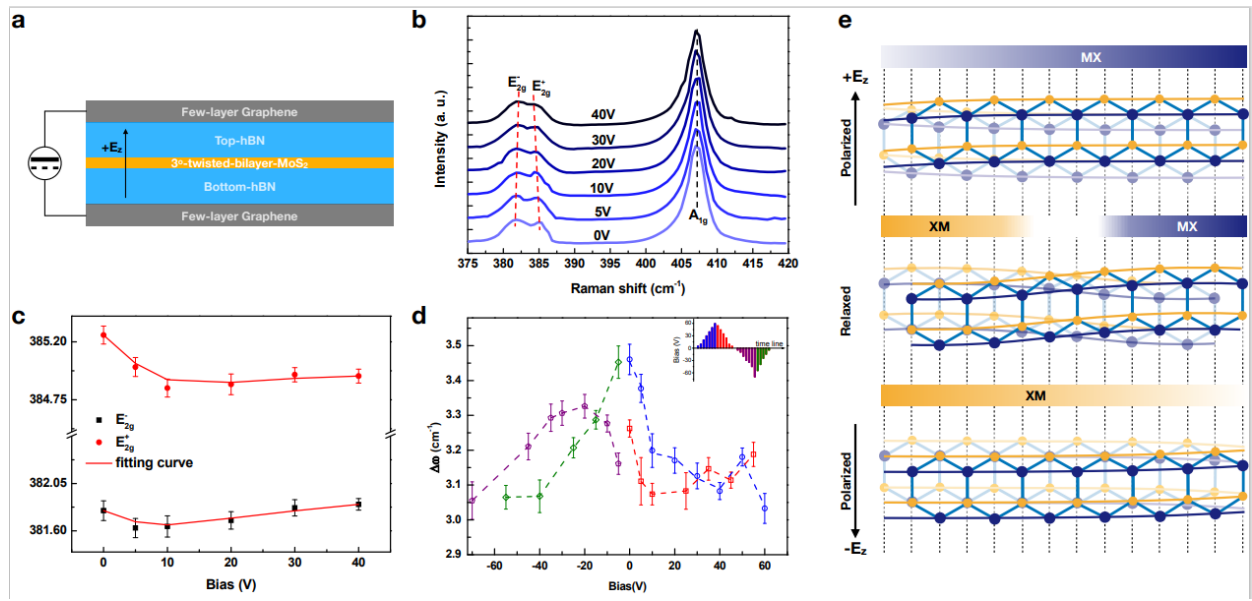

**Figure-4**

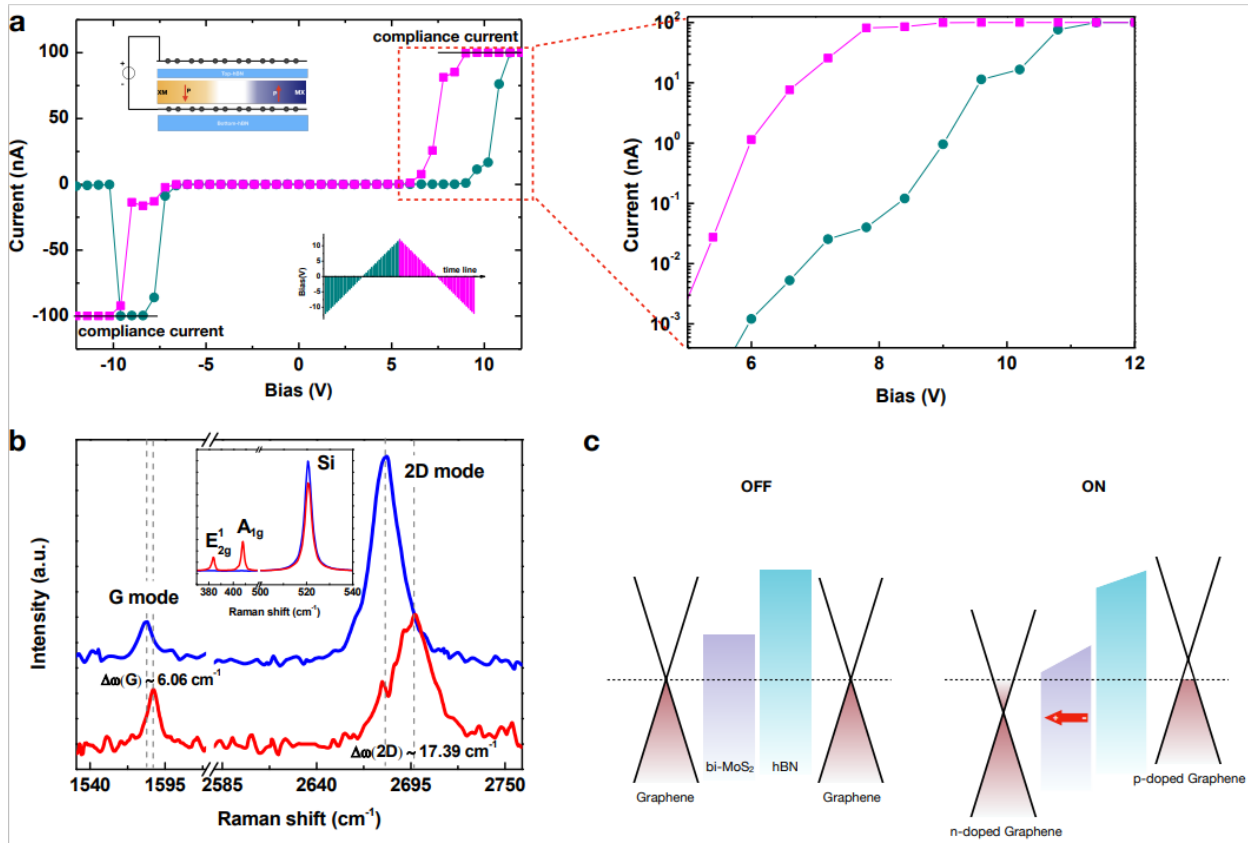